\input{aipcheck}

\documentclass[
    ,final            
  ,numberedheadings 	
  ]
  {aipproc}

\layoutstyle{6x9}

\usepackage{aas_macros}
\usepackage{natbib}
\begin{document}

\title{Calculation of the Solar UV/EUV Spectrum in Spherical Symmetry}

\author{Margit Haberreiter}{
  address={LASP, University of Colorado, 1234 Innovation Drive, Boulder, CO, 80303, USA}
}
\author{Juan Fontenla}{
}
\begin{abstract}
We present work in progress concerning spectral synthesis calculations of the solar UV/EUV in spherical symmetry carried out with the Solar Radiation Physical Modeling (SRPM) project. We compare the synthetic irradiance spectrum for the quiet Sun with the recent solar minimum spectrum taken with the EVE rocket instrument. The good agreement of the synthetic spectrum with the observation shows that the employed atmosphere structures are suitable for UV/EUV irradiance calculations. 
\end{abstract}

\maketitle

\section{Introduction}
Solar spectral irradiance variations in the UV/EUV are important for understanding and modeling the Earth's ionosphere. Here, we present work in progress toward modeling the UV/EUV variability based on spectral synthesis and feature identification from image analysis. In particular, first results of solar EUV irradiance calculations for the quiet Sun are shown. In a future step these results will be implemented in the forecasting scheme presented by Fontenla et al. \cite{Fontenla:etal:2009}. 

\section{The SRPM code}
The spectral synthesis of the UV/EUV is carried out with the Solar Radiation Physical Modeling \cite[SRPM]{Fontenla2007} project. It is a state-of-the-art radiative transfer code in full non-local thermodynamic equilibrium (NLTE). Depending on the wavelength the formation of the UV/EUV ranges from the chromosphere, transition region, and corona. Therefore, the calculations are based on semi-empirical structures that represent these layers in the solar atmosphere, as well as different features on the solar disk. For the quiet Sun these features are inter-network (IN), network lane (NL), and enhanced network (EN) \cite{Fontenla2009}. The synthetic quiet Sun spectrum is then derived as the combination of 75\% of the spectrum calculated with the model for IN, 22\% of NL, and 3\% of EN. 

For the chromosphere and transition region we solve the full NLTE for the most abundant elements from hydrogen to iron up to ion charge 2. For higher ionization states the statistical equation is solved with the optically thin approach. We account for 14,000 atomic levels and 170,000 spectral lines. 

For calculating the coronal spectrum we employ the line of sight integration in spherical symmetry \cite{Mihalas1978}, being commonly used in spectral synthesis of stars with expanding envelopes. Recently, also solar studies including the spherical setup were carried out with the COde for Solar Irradiance \citep[COSI]{Haberreiter2008b,Haberreiter2008a}. The spherical setup is essential for the EUV, as up to 50\% of the irradiance is emitted by the extended corona. 

\begin{figure}[t!]
\centering
\includegraphics[width=1.\linewidth]{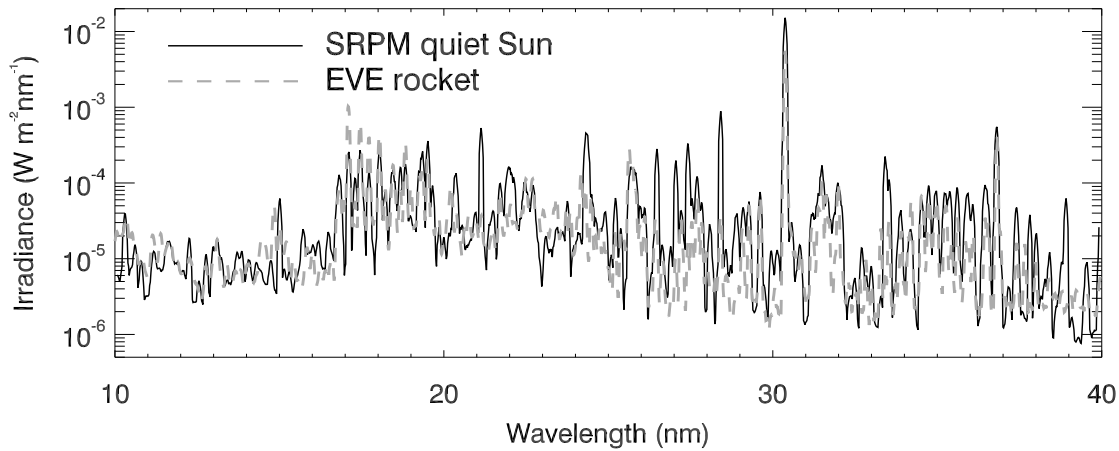}
\caption{Comparison of the synthetic quiet Sun spectrum (solid line), convolved twice with a 1-{\AA} boxcar, and the spectrum taken with the EVE rocket instrument on April 14, 2008 (gray dashed line).  \label{fig:corona}} 
\end{figure}

\section{Results and Conclusions}
As an example, Fig.\,\ref{fig:corona} shows the EUV spectrum for the quiet Sun compared with the most recent spectrum taken with the EUV Variability Experiment \cite[EVE]{Woods2006} during a rocket flight on April 14, 2008 \cite{Chamberlin2009}. As the Sun was almost free of any active region on the day of the observation, the spectrum practically represents the quiet Sun. The good agreement of the spectra shows that the employed atmosphere structures are a suitable representation of the chromosphere, transition region and corona for irradiance calculations. The results will be discussed in detail by Haberreiter and Fontenla \cite{HaberreiterFontenla2009}.
\paragraph{Acknowledgments}
This work was supported by NASA Contract NAS597045 and AFOSR Contract NNX07AO75G at the University of Colorado.

\end{document}